\documentclass[twocolumn]{aastex631}

\usepackage{hyperref,enumerate}
\usepackage{xcolor}

\hypersetup{
    unicode=false,                  
    pdftoolbar=true,                
    pdfmenubar=true,                
    pdffitwindow=true,              
    pdfstartview={FitH},            
    pdftitle={GHOST Data Reduction},
    pdfauthor={vplacco},            
    pdfsubject={Astronomy},         
    pdfcreator={dvipdf},            
    pdfproducer={dvipdf},           
    pdfkeywords={},                 
    pdfnewwindow=true,              
    colorlinks=true,                
    linkcolor=red,                  
    citecolor=blue,                 
    filecolor=magenta,              
    urlcolor=cyan,                  
    breaklinks=true,
    linktocpage
}

\begin{document}

\title{GHOST Reduced Data Products for the Gemini Observatory 
Community and Beyond}

\author[0000-0003-4479-1265]{Vinicius M.\ Placco}
\affiliation{NSF NOIRLab, 950 N. Cherry Ave., Tucson, AZ 85719, USA}

\author{David Herrera}
\affiliation{NSF NOIRLab, 950 N. Cherry Ave., Tucson, AZ 85719, USA}

\author{Brian M.\ Merino}
\affiliation{NSF NOIRLab, 950 N. Cherry Ave., Tucson, AZ 85719, USA}

\collaboration{3}{US National Gemini Office}

\author{Paul Hirst}
\affiliation{Gemini Observatory/NSF NOIRLab, 670 N. A’ohoku Place, Hilo, HI, 96720, USA}

\author[0000-0002-6633-7891]{Kathleen Labrie}
\affiliation{Gemini Observatory/NSF NOIRLab, 670 N. A’ohoku Place, Hilo, HI, 96720, USA}

\author[0000-0001-8589-4055]{Chris Simpson}
\affiliation{Gemini Observatory/NSF NOIRLab, 670 N. A’ohoku Place, Hilo, HI, 96720, USA}

\author{James Turner}
\affiliation{Gemini Observatory/NSF NOIRLab, Casilla 603, La Serena, Chile}

\author[0000-0002-9123-0068]{William D. Vacca}
\affiliation{Gemini Observatory/NSF NOIRLab, 950 N. Cherry Ave., Tucson, AZ 85719, USA}

\collaboration{5}{Gemini Science User Support Department}

\author[0000-0001-9796-2158]{Emily Deibert}
\affiliation{Gemini Observatory/NSF NOIRLab, Casilla 603, La Serena, Chile}

\author{Ruben Diaz}
\affiliation{Gemini Observatory/NSF NOIRLab, Casilla 603, La Serena, Chile}

\author[0000-0003-2530-3000]{Jeong-Eun Heo}
\affiliation{Gemini Observatory/NSF NOIRLab, Casilla 603, La Serena, Chile}

\author[0000-0002-4641-2532]{Venu Kalari}
\affiliation{Gemini Observatory/NSF NOIRLab, Casilla 603, La Serena, Chile}

\author[0000-0001-6533-6179]{Henrique Reggiani}
\affiliation{Gemini Observatory/NSF NOIRLab, Casilla 603, La Serena, Chile}

\author[0000-0002-8744-803X]{Cinthya Rodriguez}
\affiliation{Gemini Observatory/NSF NOIRLab, Casilla 603, La Serena, Chile}

\author[0000-0001-7518-1393]{Roque Ruiz-Carmona}
\affiliation{Gemini Observatory/NSF NOIRLab, Casilla 603, La Serena, Chile}

\author[0000-0003-1033-4402]{Joanna Thomas-Osip}
\affiliation{Gemini Observatory/NSF NOIRLab, Casilla 603, La Serena, Chile}

\collaboration{8}{GHOST Instrument Team}

\correspondingauthor{Vinicius M.\ Placco}
\email{vinicius.placco@noirlab.edu}

\begin{abstract}

The Gemini High-resolution Optical SpecTrograph (GHOST) at Gemini South started regular queue operations in early 2024, bringing a long-sought open-access capability to the astronomy community. This research note briefly describes an effort to provide easy-to-access reduced spectra for GHOST programs from all Gemini partner countries and encourage prompt data exploration and analysis. Since March 2024, over 4500 spectra have been reduced and made available to principal investigators (PIs). The aim is to increase demand for GHOST and expedite the publication of scientific results.

\vspace{-1.2cm}

\phantom{.}

\end{abstract}


\section{Introduction} \label{sec:intro}

The US National Gemini Office \citep[US NGO;][]{placco2023}, in collaboration with the Science User Support Department (SUSD) and the GHOST \citep{kalari2024} instrument team at the International Gemini Observatory, is leading a project to provide reduced spectra to PIs of GHOST programs (Queue, Directors Discretionary, Poor Weather, Fast Turnaround, Large and Long) from all partner countries, starting in the 2024A semester. The goals include: (i) encouraging PIs to inspect their data without installing software or running data reduction; (ii) in some cases, providing ``science-ready'' data, which can expedite analysis and publication; (iii) increasing trust regarding the Gemini data reduction software and archive; and (iv) establishing a legacy value for the archive by providing high-quality reduced spectra for the community once proprietary periods (12 months for queue programs) expire. 

\begin{figure*}[!ht]
\begin{center}
\includegraphics[width=1.000\textwidth]{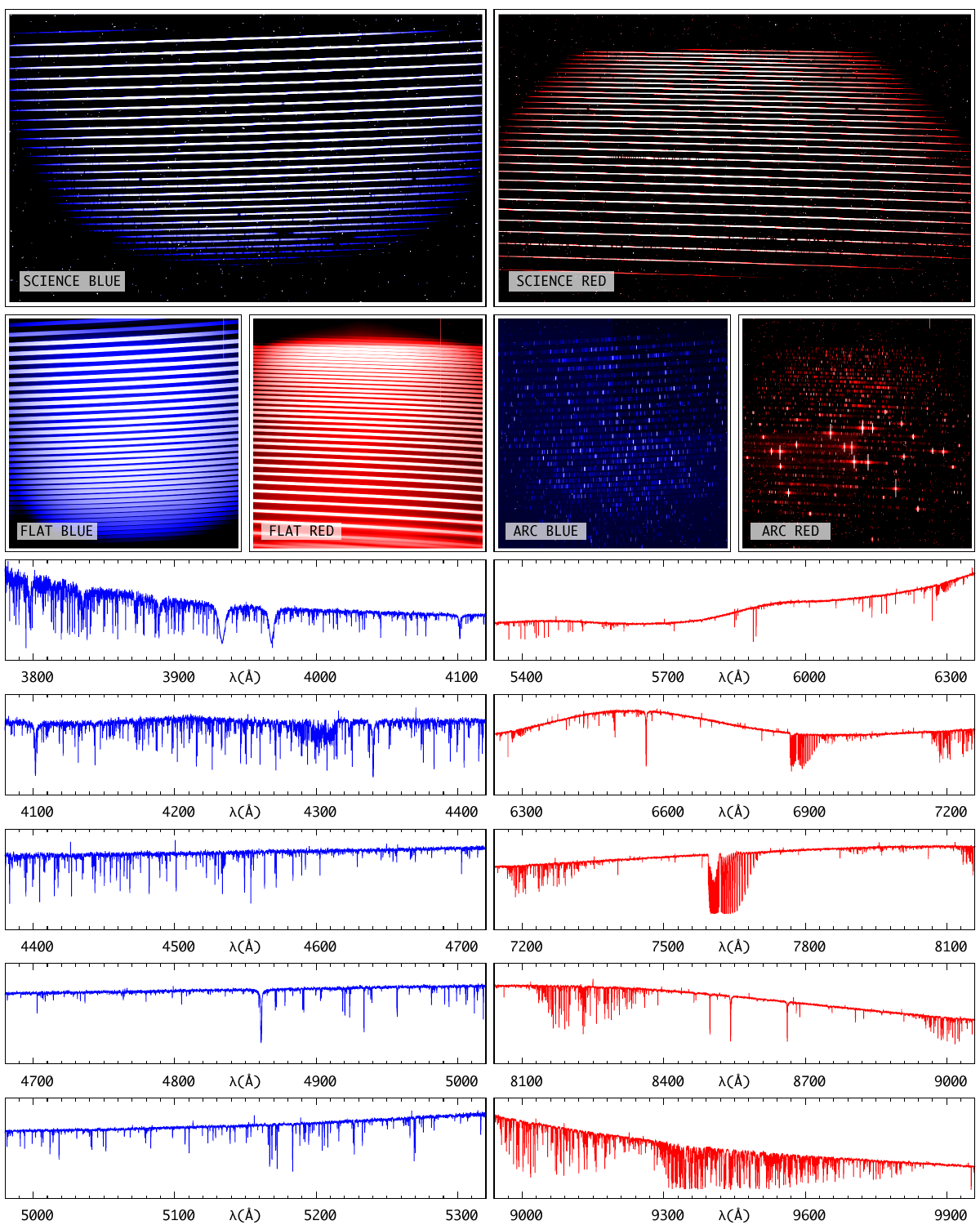}
\end{center}
\vspace{-0.5cm}
\caption{ \label{stamp} GHOST raw science (standard resolution, 2x4 binning - first row) and calibration (flat field and arc lamp - second row) data. Bottom panels: blue and red DRAGONS-reduced spectra used for quality assessment (\texttt{\_dragons.fits} files).} 
\end{figure*}

\section{Data Reduction}

The philosophy behind this project is simple: reduce every GHOST frame (with quality assessment keywords: \texttt{QA=Undefined/Usable/Pass}) tagged as \texttt{dayCal} (daytime calibrations) or \texttt{SCIENCE} that is ingested into the Gemini Observatory Archive\footnote{\href{https://archive.gemini.edu/}{https://archive.gemini.edu/}} \citep[GOA;][]{hirst2017}. The two main data reduction scripts\footnote{\href{https://gitlab.com/nsf-noirlab/csdc/usngo/ghostdr/}{https://gitlab.com/nsf-noirlab/csdc/usngo/ghostdr/}} (\texttt{calibration\_reduce.sh} and \texttt{science\_reduce.sh}) use the DRAGONS\footnote{\href{https://dragons.rtfd.io/}{https://dragons.rtfd.io/}} \citep[Data Reduction for Astronomy from Gemini Observatory North and South;][]{labrie2023} v3.2.2 \citep{simpson2024} command-line interface and run autonomously on a dedicated server hosted at NOIRLab. Calibration frames are reduced at the beginning of each night and science frames at the end of each night. All reduced \texttt{BIAS}, \texttt{FLAT}, and \texttt{ARC} frames are stored on the server, so science data can be re-reduced with different calibrations if needed.

The data reduction is performed using the default options to ensure homogeneous data products in the archive.
Every \texttt{SCIENCE} frame acquired is bias subtracted, flat corrected, wavelength calibrated, sky subtracted, and corrected for barycentric motion. Multiple exposures for the same object are not stacked and flux calibration and telluric corrections are not applied. The US NGO staff conducts a basic visual inspection and data quality assessment (see bottom panels of Figure~\ref{stamp}). 
The reduced spectra are ingested into the GOA twice a month. However, since the reduction runs daily, reduced spectra are available to PIs immediately upon request. The US NGO works with the instrument team and other National Gemini Offices to facilitate this interaction if needed, which includes checking data with varying weather conditions to assess whether a given sequence needs to be repeated. Once in the archive, reduced files become available to PIs together with the raw data for their program, following the same proprietary period rules\footnote{\href{https://www.gemini.edu/observing/policies}{https://www.gemini.edu/observing/policies}}. PIs are advised to inspect the outputs and assess whether these files can be used for scientific purposes or if non-standard reduction steps are required.

%

\vspace{-0.1cm}
\section{Current Status}

The data reduction effort started with the GHOST week-long run in late March 2024 and continued once the instrument became permanently available on April 30, 2024. The only interruptions since then were during the planned shutdown at Gemini South between July 22 and August 14, 2024, and for a week in early November for planned maintenance.
Overall, the data reduction has been extremely efficient. Between March 20th and November 30th, 97\% of the science (2095/2165 \texttt{FITS} files) and 98\% of the calibration (1711/1744 \texttt{FITS} files) data were successfully reduced. For the calibrations, most of the failures are related to minor software/hardware/header issues that were properly addressed. For the science frames, exposures considered unsuccessful include mid-exposure software failures and header issues that halt the data reduction process. Spectra where the slit viewing camera shows no flux can be reduced using a non-standard option in DRAGONS (synthetic slit view image).
In total, 9028 files were ingested into the GOA during the same period and are available to the PIs. For each raw science file, there are two reduced files for each exposure in each arm: (i) \texttt{\_calibrated.fits} are the reduced spectra before combining the multiple orders into a single red or blue spectrum; (ii) \texttt{\_dragons.fits} contain orders combined with variance weighting, with the wavelength resampled onto a log-linear scale and corrected for barycentric motion. All wavelengths are ``in-air''. DRAGONS also offers the option of outputs compatible with \texttt{IRAF} \citep{fitzpatrick2024}, \texttt{specutils.Spectrum1D}, or \texttt{ascii}.

The daily processing times for the calibration and science scripts are also being monitored. The median times are under 45 minutes for calibration, which includes 8 biases, 2 flats, and 2 arcs in the standard set (more can be added for non-standard readout modes). The science processing time depends on the number of GHOST programs available for a given set of weather conditions and RA availability. The median reduction time for the science data is around 92 minutes per night. For a few nights with many short exposures, the reduction can take up to 80 hours of non-stop processing.

\vspace{-0.20cm}
\section{Conclusion}

Providing reduced data products to PIs can be an efficient and sustainable avenue to drive demand and maximize the scientific potential of a given instrument. The feedback on the GHOST data reduction project from the community and internally at NOIRLab has been extremely positive. This effort is planned to continue for the foreseeable future (it was announced as part of the 2025A Call for Proposals) and the next step is to further advertise it within the broader astronomy community.

\vspace{-0.10cm}

%
%


\facilities{
Gemini:South (GHOST)
}

\vspace{-0.40cm}

\bibliographystyle{aasjournal}

\end{document}